\documentclass[11pt,twoside]{article}

\usepackage{asp2004}
\usepackage{epsf}
\usepackage{psfig}
\usepackage{lscape}

\newcommand{\mdot}{\ensuremath{\dot{M}}}                             

\begin{document}
\title{The impact of red giant mass loss on star cluster evolution}
\author{Jacco Th. van Loon \& Iain McDonald}
\affil{Astrophysics Group, Lennard-Jones Laboratories, Keele University,
       Staffordshire ST5 5BG, United Kingdom (jacco@astro.keele.ac.uk)}

\begin{abstract}
We discuss the importance for the long-term cluster evolution of the mass
loss from intermediate-mass stars (0.8--8 M$_\odot$). We present constraints
on the mass loss from red giants in clusters in the Magellanic Clouds, a
search for the intra-cluster medium in galactic globular clusters, and a
simple estimate for the cluster evolution due to red giant mass loss compared
to stellar escape.
\end{abstract}

\begin{figure}[!hb]
\plotfiddle{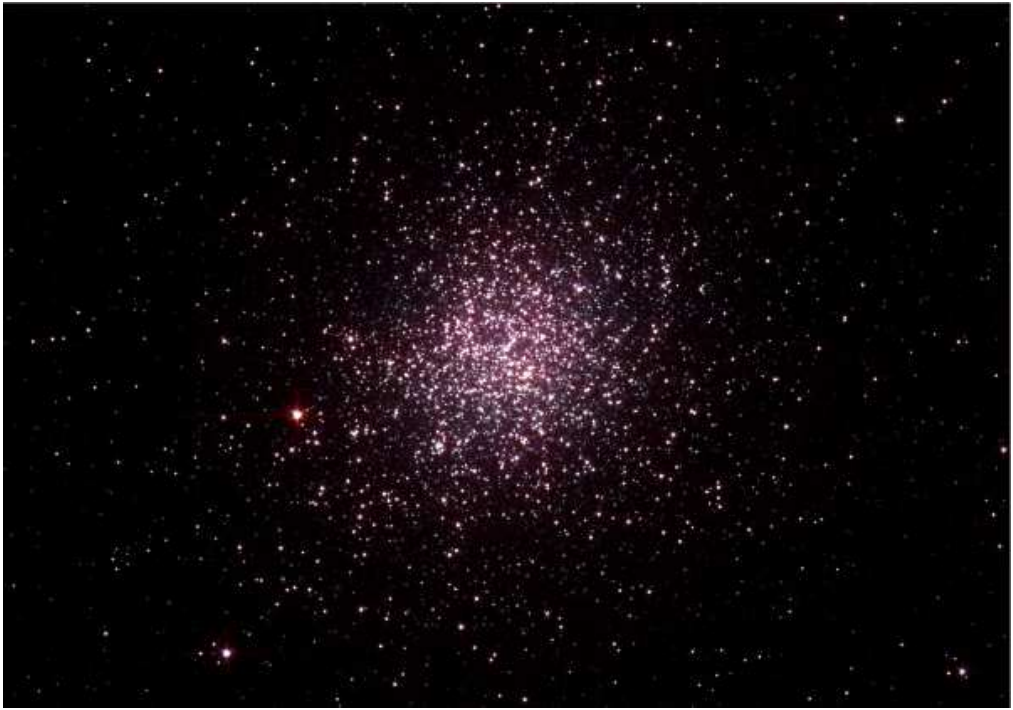}{85mm}{0}{63.3}{63.3}{-190}{-10}
\caption{Spitzer Space Telescope 3.6+4.5+5.8 $\mu$m composite of
$\omega$\,Cen.}
\end{figure}

\vspace*{-2mm}

\section{Introduction}

The evolution of a star cluster depends critically on its ability or failure
to retain its mass. Dynamical processes cause mass segregation and
preferential escape of low-mass stars from clusters. Besides losing entire
stars, clusters also lose mass due to envelope mass loss from its member
stars, exacerbating cluster dispersal. We here briefly discuss the importance
for the long-term cluster evolution of the mass loss from intermediate-mass
stars in the mass range 0.8--8 M$_\odot$.

\section{Mass loss in clusters in the Magellanic Clouds}

We searched for circumstellar dust around giants in clusters of a wide range
in ages in the SMC and LMC \citep*{vanLoonMarshallZijlstra05}. Such dusty
`superwinds' cause an infrared excess. The spectral energy distributions were
modelled to derive mass-loss rates. We compared the integrated mass-loss rate
with the total cluster mass. The timescale for a cluster to lose its mass
through dusty superwinds is found to be not much longer than the cluster age.
This is true for young clusters ($10^{7-8}$ yr) as well as older clusters
($\sim10^9$ yr). Hence, stellar mass loss can be important throughout the
evolution of a cluster.

\section{Intra-cluster medium in galactic globular clusters}

Stellar mass loss continuously replenishes the intra-cluster medium. Detecting
this material has proven difficult, suggesting efficient removal during the
journey through the galactic plane and halo. Nonetheless, we recently detected
0.3 M$_\odot$ of neutral hydrogen in the intra-cluster medium of the galactic
globular cluster M\,15 \citep{vanLoonEtal06}. Despite its low metallicity of
[Fe/H]$=-2.2$, dust has also been detected in this cluster
\citep{EvansEtal03,BoyerEtal06}.

\section{Cluster evolution as a result of stellar mass loss}

Mass loss from a star cluster can be written as
$$\mdot = c M.$$
If the mass loss is slow compared to the relaxation timescale, the process is
quasistatic. Applying the virial theorem, the cluster radius would increase as
$$R(t) \simeq R(0)\ e^{3ct}.$$
For the case of stellar escape due to cluster dynamics: $c\sim\tau_{\rm
relax}/136$. Relaxation timescales for globular clusters are typically
$\tau_{\rm relax}\sim1$ Gyr. Hence, after 10 Gyr the radius would have
increased by $\sim25$ per cent.

For a stellar initial mass function $dN \propto M^{-\alpha} dM$ with
$\alpha\sim2$, stars in the mass range 0.8--8 M$_\odot$ contribute at least as
much to the initial cluster mass as lower-mass stars. In galactic globular
clusters, with ages $t\sim10$ Gyr, these intermediate-mass stars have all lost
significant mass on the asymptotic giant branch. Assuming all this mass is
removed from the cluster, we estimate that $ct\sim0.3$. Hence, the radius will
have grown by a factor $\sim3$ over the last 10 Gyr.

We tentatively conclude that stellar mass loss could be more important for the
evolution and survival of star clusters than dynamical effects alone.

\acknowledgements We wish Henny all the best for the future.

\end{document}